\RequirePackage{ifpdf}
\documentclass[letterpaper]{JHEP3}
\usepackage{epsfig}
\usepackage{comment}
\usepackage{amsmath}

\def\ba{\begin{eqnarray}}
\def\ea{\end{eqnarray}}
\def\be{\begin{equation}}
\def\ee{\end{equation}}
\def\nn{\nonumber}
\def\exd{{\rm d}}
\def\pd{\partial}
\makeatletter
\def\x@arrow{\DOTSB\Relbar}
\def\xlongequalsignfill@{\arrowfill@\x@arrow\Relbar\x@arrow}
\newcommand{\xlongequal}[2]{%
    \ext@arrow 0099\xlongequalsignfill@{#1}{#2}}
\makeatother

\newcommand{\roughly}[1]{\mathrel{\raise.3ex\hbox{$#1$\kern-0.85em
\lower1ex\hbox{$\sim$}}}}

\def\nott#1{\setbox0=\hbox{$#1$}                
   \dimen0=\wd0                                 
   \setbox1=\hbox{/} \dimen1=\wd1               
   \ifdim\dimen0>\dimen1                        
      \rlap{\hbox to \dimen0{\hfil/\hfil}}      
      #1                                        
   \else                                        
      \rlap{\hbox to \dimen1{\hfil$#1$\hfil}}   
      /                                         
   \fi}                                         %

\def\endignore{}
\def\ignore #1\endignore{} 

\def\be{\begin{equation}}
\def\beq\begin{equation}
\def\ee{\end{equation}}
\def\bea{\begin{eqnarray}}
\def\eea{\end{eqnarray}}

\def\eqa{\begin{eqnarray}}
\def\eeqa{\end{eqnarray}}
\def\eq{\begin{equation}}
\def\eeq{\end{equation}}

\def\nn{\nonumber}

\def\pref#1{(\ref{#1})}

\def\dddot#1{#1 \hspace{-1.9mm} \ddot{\phantom{#1}} \hspace{-0.8mm} \dot{\phantom{#1}}}
\def\ddddot#1{#1 \hspace{-2.3mm} \ddot{\phantom{#1}} \hspace{-0.5mm} \ddot{\phantom{#1}}}

\def\exd{{\rm d}}
\def\nn{\nonumber}
\def\pref#1{(\ref{#1})}
\def\be{\begin{equation}}
\def\ee{\end{equation}}

\def\beq{\begin{equation}}
\def\eeq{\end{equation}}
\def\beqa{\begin{eqnarray}}
\def\eeqa{\end{eqnarray}}

\def\cD{{\cal D}}

\def\cO{{\cal O}}

\def\ssA{{\scriptscriptstyle A}}
\def\ssB{{\scriptscriptstyle B}}

\def\ssM{{\scriptscriptstyle M}}
\def\ssN{{\scriptscriptstyle N}}

\def\EFT{{\scriptscriptstyle EFT}}

\def\OPI{{\scriptscriptstyle 1\hspace{-0.3mm}PI}}
\def\OLPI{{\scriptscriptstyle 1\hspace{-0.3mm}LPI}}

\newcommand{\bmat}{\left(\begin{array}}
\newcommand{\emat}{\end{array}\right)}

\def\-{\hphantom{-}}

\def\s2{\frac{1}{2}}

\def\IF{\relax{\rm I\kern-.18em F}}
\def\II{\relax{\rm I\kern-.18em I}}
\def\IP{\relax{\rm I\kern-.18em P}}
\def\IC{\relax{\rm I\kern-.48em C}}
\def\IR{\relax{\rm I\kern-.18em R}}
\def\IK{\relax{\rm I\kern-.20em K}}
\def\IM{\relax{\rm I\kern-.25em M}}

\def\y2{Y_{\ssM\ssN} Y^{\ssM\ssN}}
\def\Riem2{R_{\ssA\ssB\ssM\ssN} R^{\ssA\ssB\ssM\ssN}}
\def\Ricci2{R_{\ssM\ssN} R^{\ssM\ssN}}

\def\f2{F^{a}_{\ssM\ssN} F^{\ssM\ssN}_a}

\def\Asl{\hbox{/\kern-.7500em\it A}} 
\def\dsl{\hbox{/\kern-.5500em$\partial$}}
\def\pxpsl{\hbox{/\kern-.5600em$p$}}
\def\Dsl{\,\raise.15ex\hbox{/}\mkern-13.5mu D}
\def \one{\relax{\rm 1\kern-.26em I}}

\def\exd{{\rm d}}

\def\nn{\nonumber}

\def\({\left(}
\def\){\right)}

\title{Who You Gonna Call?
Runaway Ghosts,\\ Higher Derivatives and Time-Dependence in EFTs}

\author{C.P.~Burgess and M.~Williams
\\
Department of Physics \& Astronomy, McMaster University,
 Hamilton ON, Canada\\
Perimeter Institute for Theoretical Physics,
 Waterloo ON, Canada\\
}

\date{}

\abstract { We briefly review the formulation of effective field theories (EFTs) in time-dependent situations, with particular attention paid to their domain of validity. Our main interest is the extent to which solutions of the EFT capture the dynamics of the full theory. For a simple model we show by explicit calculation that the low-energy action obtained from a sensible UV completion need not take the restrictive form required to obtain only second-order field equations, and we clarify why runaway solutions are nevertheless typically not a problem for the EFT. Although our results will not be surprising to many, to our knowledge they are only mentioned tangentially in the EFT literature, which (with a few exceptions) largely addresses time-independent situations.
}

\begin{document}

\section{Introduction}
\label{sec:intro}

Effective Field Theories (EFTs) are standard tools for describing situations where two very different energy scales arise, $E \ll M$, and their effectiveness is based on exploiting the simplicity that follows from expanding in powers of $E/M$ as early in a calculation as possible \cite{PhenLag, EFTrevs, EFTmine, EFTGrav}. In particular the effective (Wilson) lagrangian density is constructed exclusively from low-energy fields, but is designed to capture the virtual effects of high-energy states (with energy $M$) on the evolution of lower-energy states (with energy $E$) order-by-order in powers of $E/M$.

Implicit in this treatment is the assumption that once the high-energy modes are excluded from all initial conditions they never reappear again in final states, a property that is normally ensured by conservation of energy provided the initial energy is too small to allow transitions to the high-energy sector.\footnote{Of course, EFTs can also apply to situations where high-energy states are initially present --- such as for nucleons in the low-energy EFT for pions --- so long as they are stable (or approximately so) and so cannot catastrophically release their high energy to the lower-energy particles.} For most applications this all works because one is interested only in small fluctuations about the system's ground state, which is time-independent with only low-energy modes significantly disturbed from their vacuum.

Yet systems with different energy scales need not be prepared arbitrarily close to their vacuum, even if the energies involved are low. And more complicated states can be (and often are) time-dependent. Practical examples where this can be true include applications to cosmology or, more generally, to the response to time-dependent applied fields. How do EFTs work in such a time-dependent situation?

In this note we examine some aspects of this question, partly motivated by several recent approaches to cosmological problems. In particular we track two conceptually different (but related) issues:
 \begin{itemize}
 \item One issue works within the space of low-energy fluctuations around the vacuum and asks about how time-dependent configurations evolve within this space. In particular one asks whether solving the field equations within the relevant low-energy EFT accurately identifies the time-dependent backgrounds that would be obtained by solving the field equations of the full UV-completion.
 \item The second issue focusses on a specific time-dependent configuration identified in this way, and asks how to set up the EFT describing fluctuations about this time-dependent background (and for its domain of validity). Part of this question asks how to use conservation of energy to exclude high-energy states (as in the usual EFT development) given that fluctuations about a time-dependent background do not have a conserved energy.
 \end{itemize}
In this paper our focus is mainly on the first of these, but we argue that both issues hinge crucially on the adiabatic approximation.

For the first issue itemized above we review, in \S\ref{ssec:GenArg}, the standard argument that shows why the solutions to the EFT's equations of motion also solve the equations of motion for the full theory. Naively, this conclusion seems to lead to a problem: since EFTs generically involve interactions containing higher time derivatives, their equations of motion generically include the runaway solutions to which higher-derivative equations usually lead. How can this be true if the underlying UV completion is itself stable?

This apparent conundrum sometimes leads to the proposal that not all higher-derivative interactions actually arise in an EFT that is obtained from a stable UV completion. This proposal would be informative if true: EFTs arising from sensible underlying theories would then be subject to additional conditions beyond the usual ones of locality, cluster decomposition and so on.\footnote{Such a condition is similar in spirit to the conditions of ref.~\cite{swampland} that aim to distinguish when an EFT lies within the `landscape' of vacua of the UV theory, as opposed to the `swampland' of EFTs that do not.} In particular their higher-derivative interactions should come organized into the specific combinations that only generate second-derivative field equations, such as the Lovelock \cite{Lovelock} or Horndeski \cite{Horndeski} invariants for gravity and scalar-tensor gravity, respectively. This would be a very powerful conclusion, all the more so given that these actions (and others like them \cite{Galileon, GalileonHordeskiReview}) contain many phenomenologically interesting cosmological models \cite{Fab4,GalileonPheno}.

We show here that EFTs arising from stable UV completions in general need not be subject to an independent stability condition in this way. We first do so by explicitly computing the higher-derivative terms that arise (even at the classical level) in a simple toy model when a heavy field is integrated out. After doing so we point out how the general arguments of \S\ref{ssec:GenArg} are less general than they appear: the equations of motion of the EFT are only required to capture the effects of full theory {\em order-by-order in powers of} $1/M$, and because the runaway solutions typically vary as $e^{Mt}$ they do not arise within the $1/M$ expansion. This is why the runaway behaviour in the EFT is spurious.

Closely related to this is the observation that the EFT can only ever hope to capture the {\em adiabatic} time-dependence of the full theory, in the sense that the low-energy approximation requires time derivatives of any quantity, $\phi$, must satisfy $\dot\phi/\phi \ll M$. If this were not satisfied then generically enough energy could be extracted to invalidate the restriction to low-energy states. Although a time-dependence like $e^{\pm Mt}$ can arise within the UV completion, it would not be adiabatic and so would not be expected to be captured by the low-energy EFT \cite{CHR}.

Although we do not pursue this in detail here, we believe it is ultimately this adiabatic limit that also underlies the ability to set up an EFT describing fluctuations about a specific time-dependent configurations, such as is done for the EFT of cosmological fluctuations \cite{EFTCosmo}. Although strictly speaking the time-dependence of the background precludes the existence of a conserved energy with which to differentiate high-energy from low, for adiabatic time dependence a locally time-dependent energy {\em can} be defined for this purpose. Of course because it is time-dependent, one must continuously check that the low-energy condition, $E(t) \ll M(t)$, remains true at all times to be sure that the low-energy EFT continues to apply.

Of course none of this means there is no merit in building models from lagrangians of the Lovelock or Horndeski class, for which higher-derivative interactions are important and yet do not introduce higher than second-order field equations. Such models presuppose a regime where these particular higher derivatives are not as suppressed as are generic higher-derivative interactions. Although we do not know of examples of UV completions whose low-energy EFTs have this property, this does not mean they cannot exist and a clean enunciation of precisely when this is possible would be very instructive.

\section{The Effectiveness of the Equations of Motion}
\label{sec:EEoM}

This section presents our main results. We start, in \S\ref{ssec:GenArg}, with a review of why solutions to the EFT field equations capture the properties of solutions of the full underlying UV completion. We then specialize, in \S\ref{ssec:example}, to a simple toy model and explicitly integrate out a heavy field to verify that higher-derivative interactions are obtained that are not in the class one would consider if one were to restrict to terms that contribute only up to second derivatives in the field equations. In \S\ref{ssec:runaways} we close by showing why the arguments of \S\ref{ssec:GenArg} nonetheless do {\em not} require taking seriously the EFT's nominally runaway solutions as accurately reflecting properties of the full theory.

\subsection{General Arguments}
\label{ssec:GenArg}

To see why EFTs and UV completions agree on their solutions to the equations of motion one must hark back to the definitions of the EFT itself.\footnote{We follow here the review \cite{EFTmine}.} To this end consider a theory for which $H$ and $L$ schematically denote the `high-energy' and `low-energy' degrees of freedom, for which we wish to integrate out $H$ to obtain the EFT for $L$.

\subsubsection*{1PI Generating functionals}

A good starting point is the path-integral expression for the generator of 1PI (1-particle irreducible) correlations,\footnote{A connected graph is 1-particle reducible if it can be broken into two disconnected graphs by breaking only a single internal line.} $\Gamma(h, \ell)$, for the full theory,
\be
 \exp\Bigl\{ i \Gamma_\OPI[ h, \ell] \Bigr\} = \int \cD H \, \cD L \; \exp\left\{ i S[h + H, \ell + L] + i \int \exd^4x \, \Bigl( J H + j L \Bigr) \right\} \,,
\ee
where the `currents' $J = J(h, \ell)$ and $j = j(h, \ell)$ are implicitly defined by
\be \label{hJeq}
  \frac{\delta \Gamma_\OPI}{\delta h} + J = \frac{\delta \Gamma_\OPI}{\delta \ell} + j = 0 \,.
\ee
Although such an implicit definition at first sight might not seem very useful, it has a very simple graphical interpretation: evaluation of the currents at this point cancels the contribution of all 1-particle reducible graphs to $\Gamma_\OPI$.

In these expressions the currents $J$ and $j$ (or $h$ and $\ell$) are dummy arguments that are meant to be differentiated to obtain correlation functions, with $J = j = 0$ chosen once this differentiation is done. In particular, the field expectations, $\langle H \rangle$ and $\langle L \rangle$, for the low-energy state in which the system is prepared are given by $h$ and $\ell$  evaluated at $J = j = 0$. But this, together with eq.~\pref{hJeq}, shows that this means that these configurations are obtained by extremizing $\Gamma_\OPI$:
\be
 \left( \frac{ \delta \Gamma_\OPI}{\delta h} \right)_{h = \langle H \rangle,\, \ell = \langle L \rangle} = \left( \frac{ \delta \Gamma_\OPI}{\delta \ell} \right)_{h = \langle H \rangle,\, \ell = \langle L \rangle} = 0 \,,
\ee
and this is one of the reasons why $\Gamma_\OPI$ is of interest.

When evaluated within a semiclassical approximation we also have
\be
 \Gamma_\OPI[h, \ell] = S[h, \ell] + \Sigma_{{\rm 1-loop}}[h, \ell] + \cdots \,,
\ee
so $\Gamma_\OPI$ agrees with the classical action in the classical approximation, while $h_c = \langle H \rangle$ and $\ell_c = \langle L \rangle$ reduce to classical field configurations, that satisfy $(\delta S/\delta h)_{h_c, \ell_c} = (\delta S/\delta \ell)_{h_c, \ell_c} = 0$.

\subsubsection*{Low-energy approximation}

If only low-energy observables are of interest we can set $J = 0$ and track only $j$ (or equivalently, $\ell$). In this case it is useful to define low-energy Wilson action (or EFT) by
\be
 \exp\Bigl\{ i S_\EFT[ L ] \Bigr\} = \int \cD H  \; \exp\left\{ i S[ H, L] \right\} \,,
\ee
since this is the only part of the integral that depends on $H$. With this definition the $J=0$ result is given by
\be
 \exp\Bigl\{ i \Gamma_\OLPI[\ell] \Bigr\} = \int \cD L \; \exp\left\{ i S_\EFT[\ell + L] + i \int \exd^4x \, j L \right\} \,,
\ee
where $\Gamma_\OLPI$ denotes the generator of 1LPI (1-light-particle irreducible) correlations.
A connected graph is 1LPI if it can be broken into two disconnected graphs by breaking only one internal $L$ line, and it differs from a 1PI graph because it can include graphs that break into two when a single $H$ line is cut. $\Gamma_\OLPI$ is only 1LPI (and not 1PI) because only $j$ is evaluated at $j = - \delta \Gamma_\OLPI/\delta \ell$ to cancel the reducible graphs. $J$ can no longer similarly be used because it has been set to zero.

For later purposes what is important is that eq.~\pref{hJeq} shows that $\Gamma_\OLPI[\ell]$ is related to $\Gamma_\OPI[h, \ell]$ by
\be \label{1LPIintermsof1PI}
 \Gamma_\OLPI[\ell] = \Gamma_\OPI[ h_c(\ell), \ell] \,, \qquad \hbox{where} \quad
 \left( \frac{\delta \Gamma_\OPI}{\delta h} \right)_{h = h_c(\ell)} = 0 \,.
\ee
On the other hand, the light-field expectation, $\langle L \rangle = \ell_c$, satisfies
\be
 \left( \frac{\delta \Gamma_\OLPI}{\delta \ell} \right)_{\ell_c} = 0 \,,
\ee
which in view of eq.~\pref{1LPIintermsof1PI} and the choice made for $h = h_c(\ell)$, also shows that $\ell_c$ is also a stationary point of $\Gamma_\OPI$.

Now comes the main point. The above properties show that any configuration, $\ell_c$, obtained by extremizing $\Gamma_\OLPI$ always also extremizes $\Gamma_\OPI$, simply because $\Gamma_\OLPI$ itself is obtained from $\Gamma_\OPI$ simply by evaluating at the extremal configuration, $h = h_c(\ell)$, that satisfies $(\delta \Gamma_\OPI/\delta h)_{h = h_c} = 0$. In particular, once restricted to the classical approximation --- as is of interest in many practical applications, such as to cosmology --- the above properties show that the low-energy EFT has an action, $S_\EFT[\ell]$, that is obtained from the action, $S[h,\ell]$, of the full theory by
\be
 S_\EFT[\ell] = S[h_c(\ell), \ell] \,, \qquad \hbox{where} \quad
 \left( \frac{\delta S}{\delta h} \right)_{h = h_c(\ell)} = 0 \,.
\ee
Consequently any solution, $\ell_c$, to the field equations of the EFT,
\be
 \left( \frac{\delta S_\EFT}{\delta \ell} \right)_{\ell = \ell_c} = 0 \,,
\ee
must also be extrema of the full action, by virtue of the choice $h = h_c(\ell)$. This is why classical solutions of the effective theory are normally thought to capture the behaviour of classical solutions of the full UV-complete theory.

\subsection{An Illustrative Toy Example}
\label{ssec:example}

We now apply the above reasoning to a simple example, deriving the leading contributions to the low-energy EFT. Our goal is to show that these include higher-derivative interactions that contribute higher derivatives to the EFT's equations of motion.

Let's begin with the action for a complex scalar of the form
\be
S =-\int \! \exd^4x \Big[ \pd_\mu \phi^* \pd^\mu \phi + V(\phi^* \phi) \Big]
\ee
with
\[
 V(\phi^* \phi) = \frac{\lambda}{2} \left(\phi^* \phi - \frac{v^2}{2} \right)^2 \,.
\]
When $\lambda \ll 1$ the theory can be analyzed in a semiclassical expansion, with the vacuum obtained by minimizing $V$ at $\phi^*\phi = \frac12 \, v^2$. This spontaneously breaks the symmetry $\phi \to e^{i \omega} \phi$, leading to a particle spectrum that involves a massive field with mass $M^2 = \lambda v^2$ together with a massless Goldstone boson.

To exhibit these states explicitly we write
\[
 \phi(x) = \frac{v}{\sqrt{2}} \Bigl[ 1+ \rho(x)\Bigr] e^{i \theta(x)}
\]
where $\rho(x)$ and $\theta(x)$ are dimensionless fields. In terms of these the classical action takes the form
\be \label{ActionRTheta}
\frac{S}{v^2} = -\int \! d^4x \left[ \frac12 \pd_\mu \rho\, \pd^\mu \rho + \frac12 (1+\rho)^2 \pd_\mu \theta \,\pd^\mu \theta + V(\rho) \right]
\ee
where
\[
 V(\rho) = \frac{M^2}{2} \left(\rho^2 + \rho^3 + \frac14 \,\rho^4  \right) \,.
\]
Varying this action gives the classical equations of motion:
\begin{align} \label{EoMrho}
 \square \rho - (1+\rho) (\pd \theta)^2 - V'(\rho) &=0 \\
 \pd_\mu \Bigl[ (1+\rho)^2 \pd^\mu \theta \Bigr] &= 0 \,,
\end{align}
where $(\partial \theta)^2 := \partial_\mu \theta \partial^\mu \theta$.

In the limit where $M$ is very large compared with the energies of interest we can integrate out the $\rho$ field to determine its leading-order effects on the low-energy physics of $\theta$-particles. We do so in position space, partly to make the point that nothing in the reasoning depends on invariance under spatial translations, and so similar reasoning could be used in a gravitational context \cite{EFTGrav}. To this end we follow the above prescription and eliminate $\rho$ using the solution to its equation of motion, eq.~\pref{EoMrho}, and substitute the result back into the action, eq.~\pref{ActionRTheta}.

To obtain the solution for $\rho_c$ we introduce a function $G(x,x')$ satisfying
\be \label{GreenPDE}
(-\square + M^2)G(x,x') = \delta^{(4)}(x-x') \,,
\ee
in terms of which the (recursive) solution for $\rho(x)$ is:
\be
 \rho(x') = -\int \! \exd^4x \, G (x,x') \left\{ \Bigl[ 1+\rho(x)\Bigr] \big[ \pd\theta(x) \big]^2 + V'_{\rm int}\big[\rho(x)\big]  \right\} \label{Rrecursive}\,.
\ee
where $V_{\rm int} := V - \frac12 \, M^2 \rho^2$.

In general the solution for $\rho$ is a nonlocal mess, but simplifies considerably in the large-$M$ limit. To display this simplicity we write
\[
 \rho(x) = \sum_{n=1}^\infty \frac{r_n(x)}{M^{2n}} \,, \quad G(x,x') = \sum_{n=1}^\infty \frac{g_n(x,x')}{M^{2n}} \,,
\]
and consider only the leading and next-to-leading contributions to $\rho(x)$, up to $\cO(1/M^4)$. From eq.~\pref{GreenPDE}, we identify
\be
 g_1(x,x') = \delta^{(4)}(x-x') \,,\quad g_2(x,x') = \square \delta^{(4)}(x-x') \,,
\ee
which shows how $G(x,x')$ becomes local in the large-$M$ limit.

Using these solve for $r_1(x)$ by substituting into \pref{Rrecursive} then gives
\be
 r_1 = - (\pd\theta)^2 \,. \label{r1eval}
\ee
Interestingly, this result implies we do not require an explicit form for $r_2$ to evaluate the action to order $\cO(1/M^4)$, since using
\be
 \rho \simeq \frac{r_1}{M^2} + \frac{r_2}{M^4}
\ee
in eq.~\pref{ActionRTheta} gives
\bea \label{LEFTexplicit}
 \frac{S}{v^2} &\simeq& -\int \! \exd^4x \left\{ \frac12(\pd\theta)^2+\frac1{2M^2} \Big[ 2(\pd\theta)^2\,r_1 +r_1^2\Big] +\frac1{2M^4} \Big[(\pd r_1)^2+\big[(\pd\theta)^2 +r_1\big] \big(r_1^2+2r_2\big)\Big]\right\} \nn\\
 &=& -\int \! d^4x \left\{ \frac12(\pd\theta)^2-\frac1{2M^2}(\pd\theta)^4 +\frac1{2M^4} \pd_\mu\big[(\pd\theta)^2\big]\,\pd^\mu\big[(\pd\theta)^2\big] \right\} \\
 &=& -\int \! d^4x \left\{ \frac12(\pd\theta)^2-\frac1{2M^2}(\pd\theta)^4 +\frac2{M^4} (\theta_{\mu\nu} \theta^{\mu\rho} )\,(\pd_\rho\theta \,\pd^\nu\theta) \right\} \,,\nn
\eea
where in the last line we introduce $\theta_{\mu\nu} := \pd_\mu\pd_\nu\theta$.

To see that the last term in the action, eq.~\pref{LEFTexplicit}, potentially introduces new (often runaway) solutions it suffices to specialize to the case where all derivatives are in the time direction, in which case it is proportional to $L = \frac{k}{2} \; \ddot \theta^2 \dot \theta^2$, with $k = 4/M^4$, whose variation is
\be
 \frac{\delta L}{k} = \Bigl[\ddot \theta \dot \theta^2 \Bigr] \delta \ddot \theta + \Bigl[\dot \theta \ddot \theta^2 \Bigr] \delta \dot \theta
 = \Bigl[\ddddot \theta \dot \theta^2 +  4\, \dddot \theta \ddot \theta \; \dot \theta +  \ddot \theta^3 \Bigr] \delta  \theta  \,,
\ee
and the last equality performs several integrations by parts. Because this is a fourth-order equation for $\theta$ it requires more initial data (the initial values of $\ddot \theta$ and $\dddot \theta$), indicating the existence of new solutions.

\subsection{Clearing the runaways}
\label{ssec:runaways}

So why don't the higher-order equations of motion arising within EFTs describe solutions of the underlying UV-completion, particularly given the general arguments of \S\ref{ssec:GenArg} that appear to indicate that they should?

The key step in the previous section was the expansion in powers of $1/M$; in particular it is only after this expansion that the EFT is described by a {\em local} lagrangian density. Because of this we should only trust that integrating the equations of motion of the local EFT captures the solutions of the underlying UV-completion only order-by-order in powers of $1/M$. The problem with the `new' solutions associated with the new higher-derivative terms is that they do not arise as a series in powers of $1/M$, because they are singular perturbations of the zeroth-order differential equation.

To obtain an intuition for why this is so consider the following quadratic (but higher-order) toy lagrangian:
\be
 \frac{L}{v^2} = \frac12 \, \dot \theta^2 + \frac{1}{2M^2} \, \ddot \theta^2 \,,
\ee
whose variation $\delta L = 0$ gives the higher-order, but linear, equation of motion
\be
 - \ddot \theta  + \frac{1}{M^2} \,  \ddddot \theta  = 0 \,.
\ee
The general solution to this equation is
\be
 \theta = A + B t + C e^{Mt} + D e^{-Mt} \,,
\ee
where $A$, $B$, $C$ and $D$ are integration constants. Only the two-parameter family of these solution obtained using $C = D = 0$ go over to the solutions to the lowest-order field equation, obtained from the $M \to \infty$ lagrangian, $L_0 = \frac12 \dot \theta^2$; the other solutions are not captured at any finite order of $1/M$ because for them the $\dot\theta^2$ and $\ddot\theta^2$ terms are comparably large. Since a local EFT is only meant to capture the full theory order-by-order in $1/M$ these exponential solutions should not be expected to be relevant to the low-energy approximation of the full theory.

\section{Summary}
\label{sec:concl}

We see from this simple example why no restriction generically need be placed on the higher-derivative terms in an effective theory. In the regime where the effective theory reliably captures the behaviour of the full theory, the terms involving higher derivatives are systematically smaller than those involving fewer derivatives; a regime that does not include the worrisome runaway solutions that higher-derivative equations usually imply. The runaway solutions cannot be trusted in the regime where the effective theory must agree with the dynamics of its UV completion. 

An interesting exception to the general suppression of more derivatives in an effective theory arises in the case of the DBI action \cite{DBI}, or the action for the relativistic point particle, for which $L \sim \sqrt{1 - \dot x^2}$ can be trusted to all orders in $\dot x^2$ even while neglecting its higher derivatives, $\ddot x \simeq 0$. In this case the ultra-relativistic limit where $\dot x \to 1$ is an example of a self-consistent regime where higher derivatives are driven to zero, making it sensible to work to all orders in $\dot x$ while dropping any powers of $\ddot x$ and still-higher derivatives. (In this case symmetries also dictate how the action depends on $\dot x$, to all orders.) It would be interesting to find other examples of effective theories that share this property; theories for which all derivatives are not suppressed by the same scale and so for which it is self-consistent to consider actions that are non-trivial functions of $X = (\partial \phi)^2$ even though it is legitimate to neglect higher derivatives. It is for actions like these that restrictions on higher-derivatives might conceivably arise in interesting and constraining ways.

\section*{Acknowledgements}

We thank Andrew Tolley for useful discussions. Our research is supported in part by funds from the Natural Sciences and Engineering Research Council (NSERC) of Canada and from Perimeter Institute for Theoretical Physics. Research at the Perimeter Institute is supported in part by the Government of Canada through Industry Canada, and by the Province of Ontario through the Ministry of Research and Information (MRI).

\end{document}